\documentclass[aps,prd,twocolumn,showpacs,superscriptaddress,groupedaddress]{revtex4}  
\usepackage{graphicx}  
\usepackage{dcolumn}   
\usepackage{bm}        
\usepackage{caption}
\usepackage{amssymb}   
\usepackage{epsfig}
\usepackage{amsmath}
\usepackage{appendix}

\newcommand{\be}{\begin{equation}}
\newcommand{\ee}{\end{equation}}
\newcommand{\beq}{\begin{eqnarray}}
\newcommand{\eeq}{\end{eqnarray}}

\begin{document}

\title{Photon spheres in Einstein and 
Einstein-Gauss-Bonnet theories and circular null geodesics in axially-symmetric spacetimes}

\author{Emanuel Gallo }
 \email{egallo@famaf.unc.edu.ar}
\affiliation{ FaMAF, Universidad Nacional de C\'ordoba , Ciudad Universitaria, 5000 C\'ordoba, Argentina}
\affiliation{Instituto de F\'isica
Enrique Gaviola (IFEG) CONICET Ciudad Universitaria, 5000 C\'ordoba, Argentina.}

\author{ J. R. Villanueva }
 \email{jose.villanuevalob@uv.cl}
\affiliation{ Instituto de F\'{\i}sica y Astronom\'ia, Universidad de Valpara\'iso, 
Gran Breta\~na 1111, Valpara\'iso, Chile,}
\affiliation{Centro de Astrof\'isica de Valpara\'iso, Gran Breta\~na 1111, Playa Ancha,
Valpara\'{\i}so, Chile.}

\date{\today}

\begin{abstract}
In this article we extend 
a recent theorem proven by Hod (Phys. Lett. B, {\bf 727}, 345--348, 2013) to $n$-dimensional 
Einstein and Einstein-Gauss-Bonnet 
theories, 
which gives an upper bound for 
the photon sphere radii of spherically symmetric black holes. As  
applications of these results
we give a universal upper bound for the real part of quasinormal modes 
in the WKB limit and a 
universal lower bound for the position
of the first relativistic image in the strong lensing regime produced by these type
of black holes.
For the axially-symmetric case, we also make some general comments 
(independent of the underlying
gravitational theory) on the relation between circular null geodesics and the 
fastest way to 
circle a black hole.

\end{abstract}


\maketitle

\section{Introduction}
As is well known, the study of null geodesics at a given 
spacetime is not only 
relevant
from a theoretical point of view, but also from a practical perspective.
Their analysis, in addition to information about the causal structure of the 
spacetime, can also establish observable consequences of various 
astrophysical 
phenomena \cite{stuchlik,cov,vsoc,fernando12,vo13,vv13,garcia13}.

The role of so-called circular null geodesics is not trivial. 
In some situations, these geodesics can allow the existence of 
what is known as a photon surface, which, in the context of 
four-dimensional spacetimes, is a 
three-dimensional non-spacelike manifold $\mathcal{P}$ , such 
that every null geodesic whose
tangent vector $l^a$ at a given point $p\in \mathcal{P}$ is contained in  
the tangent space $T_p\mathcal{P}$ of $\mathcal{P}$ at $p$, always remains 
in $\mathcal{P}$. In particular, in 
the case of spherically symmetric spacetimes,
a photon sphere can be defined as a $SO(3)\times\mathcal{R}$-invariant photon surface. 
For rigorous 
definitions see\cite{Ellis01}.

The study of photon spheres in the case of spherical 
symmetry and circular null geodesics 
in axial symmetric spacetimes
 is important in astrophysics for several reasons: 
\begin{enumerate}
\item[a)] They appear explicitly in the study of 
relativistic images produced by black 
holes in the
strong lensing regime\cite{Bozza02, Bozza10}.
\item[b)] They play an important role in the analysis of quasinormal 
modes in black hole perturbations\cite{Cardoso2009, Decanini10}.
\item[c)] They determine the shadow of black holes, or 
(which is the same), how they 
look to outside observers\cite{Atamurotov13}.
\item[d)] In the case of hairy black holes, they determine 
a lower bound on the size of the hair\cite{Hod11b}.
\item[e)] They allow a link between points a) and 
b), or more precisely between gravitational waves and 
lensing, as recently shown in
\cite{Stefanov10, Wei14}.
\end{enumerate}
They have also been analyzed in situations that do not contain black holes, as 
for example, in Boson stars\cite{Horvat}, where the authors showed that 
in certain configurations 
photon spheres can occur, or in certain classes of regular metrics 
with non-negative trace of the 
energy-momentum tensor, where the existence of at least two 
photon spheres was shown\cite{Hod14}.
Recently, a study of the thermodynamics of a quantum 
version of photon spheres was presented\cite{Baldiotti14}.

Although in axially-symmetric spacetimes there is no notion yet of 
the analog of photon spheres,
there are circular null geodesics on the equatorial plane, and they are 
also useful for the
discussion of the previous points. 
For all these considerations, the characterization 
and localization of these particular surfaces or circular
null geodesics is relevant.

Recently\cite{Hod13}, Hod made some interesting observations 
about this type of null 
geodesics. In particular, in the framework of four-dimensional 
general relativity (GR), he analyzed a general family of spherically 
symmetric black holes,  which satisfy some natural asymptotic 
and energy conditions, finding 
an upper bound for the radius of photon spheres in terms of their ADM mass.
In the case of axial symmetry, he studied circular null geodesics on the 
equatorial plane of a 
Kerr spacetime, finding that they are also the fastest way to circle a 
Kerr black hole\cite{Hod11} and conjecturing an expression for the 
minimum orbital time 
to circle any compact object in GR in terms of its mass. 
As shown by Pradhan\cite{Pradhan13}, the conclusion that 
circular null geodesics are
the fastest way to circle black holes remains valid for the $n$-dimensional 
version of the Kerr-Newman metric, namely, for charged Myers-Perry spacetimes.

The reason for and interest in the study of theories of
gravity in higher dimensions is motivated by string
theory. As a possibility, the Einstein-Gauss-Bonnet (EGB) gravity
theory is selected by the low energy limit of the string
theory~\cite{cuerdas1,cuerdas3}. In this theory,
corrective terms to Einstein gravity appear, which are quadratic in the
curvature of the spacetime. The effect of those Gauss-Bonnet terms is
nontrivial for higher dimensions, so the theory of gravity, which
includes Gauss-Bonnet terms, is called Einstein-Gauss-Bonnet (EGB)
gravity. 

Even without these Gauss-Bonnet corrections,
the $n$-dimensional version of GR is usually studied in 
astrophysical and 
theoretical contexts, and in general it is also the case using 
other alternative gravitational
theories. For some works on these topics in EGB or pure GR see
\cite{Tsukamoto14, Cardoso2009, Cardoso2009b, Sadeghi14, Man14}.

Because of the evidence garnered of the importance of 
circular null geodesics
in the characterization of spherically and axially-symmetric 
black holes, 
we extend and generalize some of the Hod results to higher-dimensional 
gravitational theories. In particular we study photon spheres 
in $n$-dimensional GR and EGB theory and 
circular null geodesics on the equatorial plane for generic 
axially-symmetric spacetimes.  

The article is organized as follows. In section II we review EGB theory, 
and prepare
the setting for the discussion of photon spheres.
In section III, we establish and prove two theorems which state 
upper bounds on the radii of
photon spheres for black holes in GR and EGB theories, in 
terms of their ADM mass. 
In section IV, we analyze some of their implications, giving 
a universal upper bound for the real 
part of quasinormal frequencies in the WKB limit and a lower 
bound for the first relativistic 
image in the strong lensing regime.
In section V, we make some general comments on the relation between 
circular null geodesics in 
axially-symmetric 
spacetimes and the fastest way to circle black holes. We also show, 
by using an explicit 
counterexample, that a lower bound for the orbital period 
of circular null geodesics conjectured by Hod in the context of 
four-dimensional GR\cite{Hod11}, 
cannot be assumed to be valid also in alternative gravitational theories. 

\section{Background and setting}

The action that describes Einstein-Gauss-Bonnet gravity coupled
with matter fields reads:
\begin{equation*}
S = \frac {1}{16\pi } \int d^nx \sqrt {-g}  \left[ R - 2\Lambda +
\alpha (R_{a b c d } R^{a b c d } + \right.
\end{equation*}
\begin{equation*}
\left. + R^2 - 4R_{a b } R^{a b} )\right]+S_{\text{matt}},
\end{equation*}
where $S_{\text{matt}}$ is the action associated with the matter
fields, and $\alpha $ is the Gauss-Bonnet coupling constant
associated in the string models, with the tension of these
strings. This constant introduces a length scale. In fact, the
corrections that this theory produces to GR, are noted at short
distances, given by the scale $l=\sqrt{4 \alpha}$.

The equations of motion resulting from $\delta S=0$ are
\begin{eqnarray*}
\kappa T_{a b } &=& \mathcal{G}_{a b}=G^{(0)}_{a b}+G^{(1)}_{a
b}+G^{(2)}_{a b},
\end{eqnarray*}
where $\kappa=8\pi G/c^4$ is the gravitational constant, $T_{a b}$ 
is the energy-momentum tensor, 
representing the matter-field distribution resulting from the variation 
$\delta S_{\text{matt}}/\delta g^{a b},$ and
\begin{eqnarray*}
G^{(0)}_{a
b}&=&\Lambda g_{a b}  \\
 G^{(1)}_{a
b}&=& R_{a b }-\frac{1}{2}Rg_{a b}\\
G^{(2)}_{a b}&=&-\alpha \left[\frac {1}{2} g_{a b} (R_{ c j e
k}R^{c
 j e k }-4R_{c j }R^{c j }+R^2) \right. -   \\
&-&\left. 2RR_{a b}+4R_{a c}R^{c}_{b}+4R_{c j} R^{c j}_{ \ \ a
b}-2R_{a c j e}R_{b}^{ \ c j e} \right ].
\end{eqnarray*}

From now on, we will use unit where $c=1$ and we adopt $\alpha$ 
positive since this
condition arises from the string theory. 
We also assume asymptotically flat spacetimes, and therefore $\Lambda=0$.

Let us consider a spherically symmetric metric given by
\begin{equation}\label{eq:metric}
ds^2=-e^{-2\delta(r)}\mu(r)dt^2+\mu(r)^{-1}dr^2+r^2d\Omega^2_{n-2},
\end{equation}
with $d\Omega^2_{n-2}$
being the metric of the $(n-2)$-sphere
\begin{equation}
d\Omega^2_{n-2} = d\theta^2_1 + \sum^{n-2}_{i=2}\prod^{i-1}_{j=1}
\sin^{2}\theta_j\;d\theta^2_i \; ,
\end{equation}
solution to the Einstein-Gauss-Bonnet equations 
in higher dimensions. 
The area of the unit $(n-2)$-sphere is given by 
\begin{equation}
S_{n-2}=\frac{2\pi^{\frac{n-1}{2}}}{\Gamma\left(\frac{n-1}{2}\right)},
\end{equation} 
with $\Gamma(x)=\int^\infty_0 t^{x-1}e^{-t} dt$, the gamma function.

The Einstein-Gauss-Bonnet equations in terms of 
the components of the 
energy momentum tensor, 
$T^t_t=-\rho$, $T^r_r=p_r$ and
$T^{\theta_i}_{\theta_i}=p_\bot$ reads
\begin{equation}
\begin{split}\label{eq:rhogb}
\kappa\rho=&-\frac{1}{2r^2}(n-2)[r\mu'+(n-3)(\mu-1)]\\
&+\frac{\widehat{\alpha}}{2r^4}(n-2)(\mu-1)[2r\mu'+(n-5)(\mu-1)],
\end{split}
\end{equation}
\begin{equation}
\begin{split}\label{eq:prgb}
\kappa p_r=&-(n-2)\left\{\frac{1}{2r^2}[2r\mu\delta'-r\mu'-(n-3)(\mu-1)]\right.\\
&\left.-\frac{\widehat{\alpha}}{2r^4}(\mu-1)[-2r\mu'+4r\mu\delta'-(\mu-1)(n-5)]\right\},
\end{split}
\end{equation}
where $\widehat{\alpha}=\alpha (n-3)(n-4)$.
We have not written the angular-angular equations 
because we do not need them. Instead, 
we will use the only nontrivial component
of the energy-momentum conservation equation $\nabla_aT^a_r=0$, which reads
\begin{equation}\label{eq:conser}
p'_r=-\frac{(e^{-2\delta}\mu)'}{2e^{-2\delta}\mu}(\rho+p_r)+\frac{n-2}{r}(p_\bot-p_r).
\end{equation}
From eqs.(\ref{eq:rhogb}) and (\ref{eq:prgb}) it follows that
\begin{equation}\label{eq:delta}
\mu[2\alpha(\mu-1)-r^2]\delta'=\kappa\frac{r^3}{(n-2)}(\rho+p_r).
\end{equation}
We are interested in regular black holes.
In particular, we assume that at the horizon $r_H$, $\mu(r)$ and $\delta(r)$ satisfy
\begin{eqnarray}\label{crh}
\mu(r_H)=0,\;\;\;\;\;  \mu'(r_H)\geq 0,\\
\delta(r_H)\leq\infty,\;\;\;\;\;  \delta'(r_H)\leq\infty.
\end{eqnarray}
These conditions, together with (\ref{eq:delta}) imply
\begin{equation}
\rho(r_H)+p_r(r_H)=0.
\end{equation}

Because of the asymptotic flatness requirement, we also assume
\begin{eqnarray}
\mu(r\rightarrow \infty)=1,\label{eq:mu}\\
\delta(r\rightarrow \infty)\rightarrow 0.
\end{eqnarray}
Note that, by requiring a GR limit as $\widehat{\alpha}\rightarrow 0$,
the general solution of (\ref{eq:rhogb}) can be written as
\begin{equation}
\mu=1+\frac{r^2}{2\widehat{\alpha}}\left(1-\sqrt{1+\frac{8\kappa \widehat{\alpha} M(r)}
{(n-2)S_{n-2}r^{n-1}}}\right),\label{eq:muM}
\end{equation}
with $M(r)$ given by
\begin{equation}
M(r)=M_H+S_{n-2}\int^r_{r_H} \rho r^{n-2}dr;
\end{equation}
which can be shown to be the generalized 
Misner-Sharp mass\cite{Maeda08}. In particular,
$M_H$ is the horizon mass, and when $r$ goes 
to infinity,  $M(r)$ goes to
the ADM mass $\mathcal{M}$. In order to have a 
finite ADM mass, $\rho$ must satisfy
\begin{equation}
\lim\limits_{r\rightarrow\infty} r^{n-1}\rho=0.\label{eq:rholim}
\end{equation} 
In the vacuum case ($T^a_b=0$), we have  
\begin{equation}
\mu=1+\frac{r^2}{2\widehat{\alpha}}\left(1-\sqrt{1+\frac{8\kappa \widehat{\alpha}\mathcal{M}}
{(n-2)S_{n-2}r^{n-1}}}\right);
\end{equation}
and the associated vacuum metric is known as the 
Boulware-Deser-Wheeler (BDW) black hole.

If $\hat{\alpha}\rightarrow 0$, Eq.(\ref{eq:muM}) reduces to its GR 
limit,
\begin{equation}
\mu=1-\frac{2\kappa M(r)}
{(n-2)S_{n-2}r^{n-3}}.\label{eq:muMe}
\end{equation}

Finally, 
\begin{equation}
T=-\rho+p_r+(n-2)p_\bot;
\end{equation} 
denotes the trace of the energy momentum tensor and we assume 
that matter satisfies the dominant energy condition (DEC) 
$\rho\geq 0$, $\rho\geq |p_r|,|p_\bot|$. From
the DEC, we see then that $M_H\leq M(r)\leq \mathcal{M}.$

\section{Upper bound on the photon sphere radii}
We are now prepared to establish the following theorems:\\

\textbf{Theorem 1. Hod's theorem ($n$-dimensional GR version):} \textit{Let $(\tilde{M},g_{ab})$ 
be a spherically symmetric spacetime, such that: i) it has a regular 
event horizon, ii) it is asymptotically flat, iii) the dominant 
energy condition 
is satisfied and the 
energy-momentum trace is nonpositive, iv) it satisfies the $n$-dimensional 
Einstein equations. 
Then this metric admits at least one photon sphere, which in the coordinates 
given by (\ref{eq:metric}) is characterized by
a radius $r_\gamma$, which is bounded by the
following expression in terms of its total ADM mass $\mathcal{M}$:
\begin{equation}
r_\gamma\leq\left(\frac{\kappa(n-1)}{(n-2)S_{n-2}} \mathcal{M}\right)^{\frac{1}{n-3}}.\label{boundrgr}
\end{equation}
}

This theorem can be extended to Einstein-Gauss-Bonnet theory 
in $n$-dimensions. In particular, 
in five dimensions we can obtain an upper bound for the photon sphere radius in terms of 
the ADM mass and the constant $\hat\alpha$.
For $n>5$, although we cannot give a general expression for the 
photon sphere radius bound in terms of the ADM mass 
(with the exception of $n=9$), 
we can state a more general result. These properties 
are summarized in the following theorem.\\

\textbf{Theorem 2. $n$-dimensional EGB version:} \textit{Let $(\tilde{M},g_{ab})$ be a 
spherically symmetric spacetime such that: a) it satisfies conditions i), ii) 
and iii) of Theorem $1$, b) it satisfies the $n$-dimensional 
EGB equations. Then this metric admits at least one photon sphere, 
whose radius $r_\gamma$ in the coordinates given by (\ref{eq:metric}) is always 
bounded
by the photon sphere radius $r^*_\gamma$ of a vacuum spherically 
symmetric BDW black hole 
with the same total ADM mass $\mathcal{M}$, i.e. $r_\gamma\leq r^*_\gamma$.
In particular in five dimensions the photon sphere radius is bounded by the
following expression in terms of its total ADM mass $\mathcal{M}$:
\begin{equation}
r_\gamma\leq\frac{\sqrt{6}[\kappa\mathcal{M}
(\kappa\mathcal{M}-3\widehat{\alpha}\pi^2)]^{1/4}}{3\pi}.\label{rg5}
\end{equation}
}
 
Note that the inequality (\ref{rg5}) is saturated for a five-dimensional BDW 
black hole with mass
$\mathcal{M}$.\\

\textbf{Proof:} \textit{Common part.} Since the Einstein equations 
can be recovered from (\ref{eq:rhogb}) and (\ref{eq:prgb}) with 
$\hat{\alpha}=0$, in the proof 
there is a common part for both theorems. 
Basically, we follow the same steps as Hod's proof\cite{Hod13}, although some 
changes are 
needed to incorporate the $n$-dimensional case. 
Because of the spherical symmetry, null geodesics may without 
loss of generality be taken to be 
on the equatorial plane. 
From the metric (\ref{eq:metric}) it is easy to show that the equation
for circular null geodesics $\dot r_\gamma=(\dot r_\gamma)'=0$ 
(where a dot means derivative with
respect to an affine parameter and $'$ means derivative with respect to $r$)
reads\cite{Chandra83,Cardoso2009}
\begin{equation}
\tilde{N}(r_\gamma)=0;
\end{equation}
with
\begin{equation}
\tilde{N}(r)=2e^{-2\delta(r)}\mu(r)-r[e^{-2\delta(r)}\mu(r)]'.
\end{equation}
By defining the function
\begin{equation}
N(r)=e^{2\delta(r)}\tilde{N}(r),\label{eq:No}
\end{equation}
we similarly obtain
\begin{equation}\label{N1}
N(r_\gamma)=2[1+r_\gamma\delta'(r_\gamma)]\mu(r_\gamma)-r_\gamma\mu'(r_\gamma)=0.
\end{equation}

From the trace of the energy-momentum tensor and (\ref{eq:No}),
we can write (\ref{eq:conser}) as
\begin{equation}
p'_r=\frac{1}{2\mu r}\left(N(\rho+p_r)+2\mu T-2n\mu p_r\right).\label{eq:prp}
\end{equation}
Now if we define:
\begin{equation}
P=r^np_r,
\end{equation}
we arrive at
\begin{equation}\label{consef}
P'(r)=\frac{r^{n-1}}{2\mu}\left(N(\rho+p_r)+2\mu T\right). 
\end{equation}
In terms of the energy-momentum components, $N$ reads
\begin{equation}\label{N2}
\begin{split}
N=&\frac{1}{2\widehat{\alpha}(\mu-1)-r^2}\left\{(n-1)\widehat{\alpha}\mu^2\right.\\
&\left.-[(n-1)r^2+2(n-3)\widehat{\alpha}]\mu+(n-5)\widehat{\alpha}+(n-3)r^2\right.\\
&\left.+\frac{2}{n-2}\kappa r^4p_r\right\};
\end{split}
\end{equation}
which in the $n$-dimensional GR
case reduces to
\begin{equation}
N_{GR}=(n-1)\mu-(n-3)-\frac{2\kappa}{(n-2)}r^2p_r.\label{NGR}
\end{equation}
Let us observe that (\ref{N1}) admits at least one solution as follows 
from the fact that 
the conditions (\ref{crh}) and (\ref{eq:No}) imply 
$N(r_H)\leq 0$, and taking into account that 
$\lim\limits_{r\rightarrow\infty} r^2p_r=0$ (which follows 
from the dominant energy condition and (\ref{eq:rholim})), we see that
$N(r\rightarrow\infty)\rightarrow 2$; therefore, there 
is a $r_\gamma$ where $N(r_\gamma)=0$.
What is more, because we are interested in the innermost 
null circular orbit, $N(r)$
 must satisfy 
\begin{equation}\label{N3}
N(r_H\leq r<r_\gamma)<0.
\end{equation}
Then, by using the dominant energy condition 
we obtain
\begin{equation}
p_r(r_H)=-\rho(r_H)\leq 0, \;\; \Rightarrow p_r(r_H)\leq 0 \Leftrightarrow P(r_H)\leq 0.\label{prh1}
\end{equation}
Similarly from (\ref{consef}), (\ref{N3}) and the condition of nonpositive trace $T\leq 0$ , 
we see that
 $P'(r_H\leq r<r_\gamma)\leq 0.$ 
From this last condition and (\ref{prh1}) we have
\begin{equation}
p_r(r_\gamma)\leq 0.\label{eq:prf}
\end{equation}
Let us analyze the implications of this relation in the Einstein 
and Einstein-Gauss-Bonnet cases
separately.\\

{\textsl{ Completion of the proof of Theorem 1.}}
In the Einstein case, from (\ref{eq:prf}) and (\ref{NGR}) we conclude that
\begin{equation}
\mu(r_\gamma)\leq\frac{n-3}{n-1},
\end{equation}
which  by using (\ref{eq:muMe}) implies
\begin{equation}
r_\gamma\leq\left(\frac{\kappa(n-1)}{(n-2)S_{n-2}}{M(r_\gamma)}\right)^{\frac{1}{n-3}},
\end{equation}
or in terms of the total ADM mass $\mathcal{M}$,
\begin{equation}
r_\gamma\leq\left(\frac{\kappa(n-1)}{(n-2)S_{n-2}} \mathcal{M}\right)^{\frac{1}{n-3}}.
\end{equation}

\textsl{Completion of the proof of Theorem 2.} Let us start by 
noting that in (\ref{N2}) 
the following expression appears in 
the denominator
\begin{equation}
2\widehat{\alpha}(\mu-1)-r^2=-\frac{r^2}{2\widehat{\alpha}}\sqrt{1
+\frac{8\kappa \widehat{\alpha} 
M(r)}{(n-2)S_{n-2}r^{n-1}}}<0.
\end{equation} 
Therefore from (\ref{N2}) and (\ref{eq:prf}) we deduce that at $r_\gamma$,

\begin{equation}\label{Ngb}
\begin{split}
&N^*\equiv\left\{(n-1)\widehat{\alpha}\mu^2-[(n-1)r^2+2(n-3)\widehat{\alpha}]\mu
+(n-5)\widehat{\alpha}
\right.\\
&\left.+(n-3)r^2\right\}=-\frac{2}{n-2}\kappa r^4p_r\geq0.
\end{split}
\end{equation}
The implications of this equation for the allowed $r_\gamma$ 
will be discussed 
first in five dimensions because this case is exactly solvable, 
and after that, we will
extend the analysis to higher dimensions.

By writing $N^*$  
as
\begin{equation}
N^*=(n-1)\widehat{\alpha}(\mu-B_{(n)-})(\mu-B_{(n)+}),\label{eq:NbmbM}
\end{equation}
we see that in five dimensions, (\ref{Ngb}) implies that at $r_\gamma$, some
of these conditions hold:
 \begin{equation}\label{eq:mup}
\mu\leq\frac{1}{2\widehat{\alpha}}(\widehat{\alpha}+r^2
-\sqrt{\widehat{\alpha}^2+r^4})\equiv B_{(5)-},
\end{equation}

or
\begin{equation}\label{eq:munp}
\mu\geq\frac{1}{2\widehat{\alpha}}(\widehat{\alpha}+r^2+\sqrt{\widehat{\alpha}^2+r^4})\equiv B_{(5)+};
\end{equation}
where in order to write these inequalities we use the fact 
that $B_{(5)-}$ and $B_{(5)+}$ are both 
positive real 
functions and that $B_{(5)-}<B_{(5)+}$.

We also note [using $M_H\leq M(r)\leq \mathcal{M}$] that independently of 
the dimension,
$\mu$ is bounded from below by
\begin{equation}\label{eq:muMadm}
\mu\geq 1+\frac{r^2}{2\widehat{\alpha}}\left(1-\sqrt{1+\frac{8\kappa \widehat{\alpha} \mathcal{M}}
{(n-2)S_{n-2}r^{n-1}}}\right)\equiv \mu_{(n)\mathcal{M}};
\end{equation}
and by 
\begin{equation}\label{eq:muMH}
\mu\leq 1+\frac{r^2}{2\widehat{\alpha}}\left(1-\sqrt{1+\frac{8\kappa \widehat{\alpha} {M_H}}
{(n-2)S_{n-2}r^{n-1}}}\right)\equiv \mu_{(n){M_H}};
\end{equation}
from above.
Consequently, the validity of the conditions (\ref{eq:mup}) or (\ref{eq:munp}) implies
\begin{equation}
\mu_{(5)\mathcal{M}}(r_\gamma)\leq B_{(5)-}(r_\gamma);\label{in1}
\end{equation}
or 
\begin{equation}
\mu_{(5)M_H}(r_\gamma)\geq B_{(5)+}(r_\gamma);\label{in2}
\end{equation}
respectively.
Moreover, the four functions $\mu_{\mathcal{M}}(r)$, $\mu_{(5)M_H}$, $B_{(5)+}(r)$ and 
$B_{(5)-}(r)$ are 
all monotonically increasing functions.
From this fact, and observing that $B_{(5)+}(r)\geq 1$, and $\mu_{(5)M_H}(r\rightarrow\infty)=1$, 
it follows that 
\begin{equation}
\mu(r)\leq\mu_{(5)M_H}(r) < B_{(5)+}(r)\; \forall r,
\end{equation} 
thus,  
it is not possible for any real $r_\gamma$ to satisfy the 
inequality (\ref{eq:munp}). 

Let us study now the inequality (\ref{in1}). 
Observing that for $r\geq r_H$,
\begin{equation}
0<B_{(5)-}(r)<B_{(5)-}(r\rightarrow\infty)=\frac{1}{2},
\end{equation}
and 
\begin{equation}
\mu_{(5)\mathcal{M}}(r=r_H)\leq 0,\;\; \mu_{(5)\mathcal{M}}(r\rightarrow\infty)=1,
\end{equation} 
we see that (\ref{in1}) will be valid for all $r$ such that 
$r_H\leq\, r\leq r^*_\gamma$, with
$r^*_\gamma$ solution of 
 \begin{equation}
\mu_{(5)\mathcal{M}}(r^*_\gamma)=B_{(5)-}(r^*_\gamma).
\end{equation}
The only positive real solution of this equation is 
\begin{equation}
r^*_\gamma=\frac{\sqrt{6}[\kappa\mathcal{M}(\kappa\mathcal{M}-3\widehat{\alpha}\pi^2)]^{1/4}}{3\pi}.
\end{equation}
On the other hand, as $\mu(r_H)=0$, $B_{(5)-}(r_H)>0$, and $\mu(r)\geq \mu_{(5)\mathcal{M}}(r)$, 
we conclude that  $\mu(r)> B_{(5)-}(r)$ for all $r> r^*_\gamma$.
Consequently, the radius $r_\gamma$ which satisfies (\ref{eq:mup}) 
must necessarily be
bounded from above by $r^*_\gamma$, that is
\begin{equation}
r_\gamma\leq\frac{\sqrt{6}[\kappa\mathcal{M}(\kappa\mathcal{M}-
3\widehat{\alpha}\pi^2)]^{1/4}}{3\pi}.\label{eqrg}
\end{equation}

In the $n$-dimensional case,  as was mentioned above, 
we cannot find an explicit upper bound in terms of the ADM mass;
however, we can show that even in these situations, the 
photon sphere radius will always 
be bounded by the photon sphere radius of a BDW black 
hole with the same mass.
It can be shown as follows. 

First, in the general case, following a similar analysis 
as in Eqs.(\ref{eq:NbmbM}), (\ref{eq:mup}) 
and (\ref{eq:munp}), we see that (\ref{Ngb}) is satisfied at $r_\gamma$ if
\begin{equation}\label{eq:mupN}
\mu\leq\frac{1}{2\widehat{\alpha}}\left(\frac{2(n-3)}{n-1}\widehat{\alpha}+r^2-
\sqrt{\frac{16}{(n-1)^2}\widehat{\alpha}^2
+r^4}\right)\equiv B_{(n)-},
\end{equation}
or if 
\begin{equation}\label{eq:mupN2}
\mu\geq\frac{1}{2\widehat{\alpha}}\left(\frac{2(n-3)}{n-1}\widehat{\alpha}+r^2+
\sqrt{\frac{16}{(n-1)^2}\widehat{\alpha}^2
+r^4}\right)\equiv B_{(n)+},
\end{equation}
which using (\ref{eq:muMadm}) and (\ref{eq:muMH}) implies
\begin{equation}
\mu_{(n)\mathcal{M}}(r_\gamma)\leq B_{(n)-}(r_\gamma);\label{inn1}
\end{equation}
or 
\begin{equation}
\mu_{(n)M_H}(r_\gamma)\geq B_{(n)+}(r_\gamma).\label{inn2}
\end{equation}
For the same reasons as in the five-dimensional case [$B_{(n)+}(r)\geq 1$, 
$\mu_{(n){M_H}}(r\rightarrow\infty)=1$, and both being 
monotonically increasing functions], 
we conclude that there is no real $r_\gamma$ so that (\ref{inn2}) [and hence 
(\ref{eq:mupN2})] 
can be satisfied.
On the other hand, from the fact that $B_{(n)-}$ and $\mu_{(n)\mathcal{M}}$ are
monotonically increasing functions, 
and taking into account that
\begin{equation}
0\leq B_{(n)-}(r)< B_{(n)-}(r\rightarrow\infty)=\frac{n-3}{n-1}<1\;\forall r\geq r_H;
\end{equation} and $\mu_{(n)\mathcal{M}}(r_H)\leq 0$, 
$\mu_{(n)\mathcal{M}}(r\rightarrow\infty)=1$, we see that (\ref{inn1}) holds
for all $r_H\leq r<r^*_\gamma$ with $r^*_\gamma$ satisfying
\begin{equation}
\mu_{(n)\mathcal{M}}(r^*_\gamma)=B_{(n)-}(r^*_\gamma).\label{eq:genAB}
\end{equation}
As 
$\mu(r_H)=0, B_{(n)-}(r_H)>0$ and $\mu(r)\geq \mu_{(n)\mathcal{M}}(r)$; and following the
same arguments as in the five-dimensional situation, we conclude that  
the radius $r_\gamma$ where (\ref{eq:mup}) holds, will also
satisfy $r_\gamma\leq r^*_\gamma$.
Furthermore, from (\ref{N2}), (\ref{N3}) and (\ref{eq:NbmbM}), and observing that
\begin{equation}
N^*=[2\widehat{\alpha}(\mu-1)-r^2]N-\frac{2}{n-2}\kappa r^4p_r,
\end{equation} 
we see that (\ref{eq:genAB}) is equivalent 
to requiring $N(r^*_\gamma)=0$ for a 
BDW black hole with ADM-mass $\mathcal{M}$ ($\rho=p_r=p_\bot=0$).
In other words, $r^*_\gamma$ can be interpreted as the radius 
of the photon sphere for a 
BDW  black hole with mass $\mathcal{M}$. As a result, 
the bound is 
saturated in these cases. $\blacksquare$\\

Let us make some remarks before continuing. In order to see how 
the bound (\ref{boundrgr}) varies 
in terms of the dimension, we plot the quotient $r_\gamma/(G\mathcal{M})^{\frac{1}{n-3}}$ 
in Fig.(1), assuming a 
continuous $n$. The physical values 
must be taken
only for natural $n$ in the graphic. We see that in eight dimensions 
this quotient reaches its 
minimum value. However, as for $n>4$, the dependence in the mass is not linear, 
the dimension which 
minimizes the value of the upper bound
for the radius $r_\gamma$ at a given fixed $\mathcal{M}$ will
depend on the value of this mass.
\begin{figure}[!h]
\centering 
\includegraphics[clip,width=80mm]{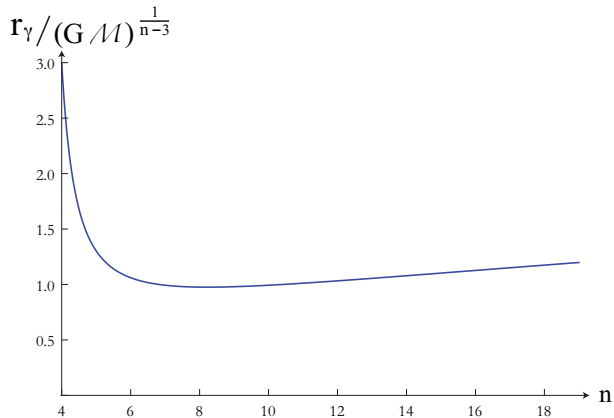}
\caption{Upper bound for the quotient $r_\gamma/(G\mathcal{M})^{\frac{1}{n-3}}$ 
in terms of the 
dimension $n$ of the spacetime.
}
\label{fig:distrib}
\end{figure}

Let us also observe that if we take $\alpha\rightarrow 0$ in (\ref{eqrg}), 
we recover the upper 
bound that we found in the GR 
case.

As a final comment, we mention that (\ref{eq:genAB}) can also be 
solved in nine dimensions, obtaining 
the result
\begin{equation}
r_\gamma\leq 98(14)^{1/3}\left[\frac{h^{2/3}-686(14)^{2/3}\pi^2\kappa\hat\alpha\mathcal{M}
}{\pi^3h^{1/3}}\right]^{1/4},
\end{equation}
with
\begin{equation}
h=\pi\kappa\mathcal{M}\left[18\kappa\mathcal{M}+\sqrt{2\kappa\mathcal{M}
(162\kappa\mathcal{M}+7\pi^4\hat\alpha^3)}\right].
\end{equation}

\section{Applications: Quasinormal modes and strong lensing}

Recently(\cite{Cardoso2009}), Cardoso {\it et al.}, showed a 
correspondence between
the quasinormal modes associated to a black hole,
in the eikonal limit, and some properties 
of photon spheres.
In particular, in this limit, the quasinormal
frequencies $\omega_{QNM}$ can be computed from 
analytical WKB 
approximation methods\cite{Iyer87},
obtaining
\begin{equation}
\omega_{QNM}=l\Omega_\infty-i|\lambda|(m+1/2),\label{eq:wqnm}
\end{equation}
where 
\begin{equation}
\Omega_\infty=\frac{[e^{-2\delta(r_\gamma)}\mu(r_\gamma)]^{1/2}}{r_\gamma},\label{omegainfty}
\end{equation} 
is the angular velocity
of circular null geodesics at $r_\gamma$ as
measured by asymptotic observers, $\lambda$ is the 
Lyapunov exponent associated with
this kind of geodesics, 
$l$ is the angular momentum of the perturbation 
(and it is assumed that $l\gg 1$) 
and $m$ is the overtone number. 
Based on this fact, Hod in\cite{Hod13} presented a simple 
and universal bound
for the real part of the quasinormal frequencies
in the WKB limit, expressed in terms of the event horizon 
radius of the associated black hole.
Now, we generalize this expression to $n$-dimensional 
GR and EGB theory, and we also
discuss other simple bounds for other observables. 

From (\ref{eq:wqnm}), we see that the real part of these frequencies is given by
\begin{equation}
\omega_l\equiv\Re{[\omega_{QNM}]}=l\Omega_\infty.\label{omegalq}
\end{equation}
By using (\ref{eq:mupN}), which can be re-written as
\begin{equation}
\mu(r_\gamma)\leq 2\frac{\frac{n-5}{n-1}\hat\alpha+\frac{n-3}{n-1}r^2_\gamma}
{\frac{n-3}{n-1}\alpha+r^2_\gamma+\sqrt{\frac{16\hat\alpha^2}{(n-1)^2}+r^4_\gamma}};
\end{equation}
and using the fact that $e^{-2\delta}\leq 1$, [which follows from (\ref{eq:delta}) and 
the asymptotic 
condition for $\delta(r)$], we see that (\ref{omegainfty}) implies 
\begin{equation}
\Omega_\infty\leq 2\left[\frac{\frac{n-5}{(n-1)r^2_\gamma}\hat\alpha+\frac{n-3}{n-1}}
{\frac{n-3}{n-1}\alpha+r^2_\gamma+\sqrt{\frac{16\hat\alpha^2}{(n-1)^2}+r^4_\gamma}}\right]^{1/2}.
\end{equation}
Hence, from relation (\ref{omegalq}) we obtain
\begin{equation}\label{eq:omegalp}
{\omega_l}\leq 2 l \left[\frac{\frac{n-5}{(n-1)r^2_\gamma}\hat\alpha+\frac{n-3}{n-1}}
{\frac{n-3}{n-1}\alpha+r^2_\gamma+\sqrt{\frac{16\hat\alpha^2}{(n-1)^2}+r^4_\gamma}}\right]^{1/2}.
\end{equation}
In order to compare this with the Hod bound, we will give a 
weaker inequality, but one 
which is simpler and more universal than (\ref{eq:omegalp})
in the sense that it does not depend on the 
particular properties of the
black hole.
Let us note that, in our study cases, $\hat\alpha\geq 0$ and $r_\gamma\geq r_H$; 
therefore,

\begin{equation}
\begin{split}
{\Omega}_\infty & \leq 2  \left[\frac{\frac{n-5}{(n-1)r^2_\gamma}\hat\alpha+\frac{n-3}{n-1}}
{\frac{n-3}{n-1}\hat\alpha+r^2_\gamma+\sqrt{\frac{16\hat\alpha^2}{(n-1)^2}+r^4_\gamma}}\right]^{1/2}\\
& \leq 2  \left[\frac{\frac{n-5}{(n-1)r^2_H}\hat\alpha+\frac{n-3}{n-1}}
{\frac{n-3}{n-1}\hat\alpha+r^2_H+\sqrt{\frac{16\hat\alpha^2}{(n-1)^2}+r^4_H}}\right]^{1/2}\\
& \leq   \left[\frac{\frac{n-5}{(n-1)r^2_H}\hat\alpha+\frac{n-3}{n-1}}
{r^2_H}\right]^{1/2}.\label{omegacot}
\end{split}
\end{equation}

Combining this inequality with (\ref{omegalq}), we finally obtain
 \begin{equation}
\begin{split}
{\omega}_l 
\leq l  \left[\frac{\frac{n-5}{(n-1)r^2_H}\hat\alpha+\frac{n-3}{n-1}}
{r^2_H}\right]^{1/2};
\end{split}
\end{equation}
which implies
 \begin{equation}
\begin{split}
{\omega}_l r_H 
& \leq l  \left[\frac{n-5}{(n-1)r^2_H}\hat\alpha+\frac{n-3}{n-1}\right]^{1/2}.\label{eq:omeg}
\end{split}
\end{equation}
This expression generalizes to $n$-dimensional EGB theory, 
the universal bound found
by Hod in the framework of four-dimensional GR\cite{Hod13}.
In particular, in the GR case it reduces to 
\begin{equation}
{\omega}_l r_H  \leq l  \left[\frac{n-3}{n-1}\right]^{1/2},
\end{equation}
which, when $n=4$ agrees with the expression given in \cite{Hod13}.
Note also that in five dimensions the first factor of 
(\ref{eq:omeg}) does not survive,  
and we get the simple bound
\begin{equation}
{\omega}_l r_H \leq \frac{1}{\sqrt{2}} l.
\end{equation}

At this point, we can ask if there is a similar 
universal bound for the 
Lyapunov exponent; 
however, the answer is negative.
The Lyapunov exponent for circular null geodesics 
is defined by\cite{Cardoso2009}
\begin{equation}
\lambda=\sqrt{\frac{r^2_\gamma [e^{-2\delta(r_\gamma)}\mu(r_\gamma)]}
{2L^2}V''(r_\gamma)},\label{lambda}
\end{equation}
with 
\begin{equation}
V''(r_\gamma)=\frac{L^2}{r^4_\gamma e^{-2\delta{(r_\gamma)}}}\left\{e^{-2\delta{(r_\gamma)}}
\mu(r_\gamma)-r^2_\gamma [e^{-2\delta(r_\gamma)}\mu(r_\gamma)]''\right\},
\end{equation}
and $L$ the orbital angular momentum of the circular null geodesics.
After some simple computations it can be shown that in $n$-dimensional GR,
\begin{equation}
V''(r_\gamma)=\frac{L^2 N'(r_\gamma)}{r^3_\gamma}.
\end{equation}
This makes it possible to write (\ref{lambda}) as
\begin{equation}
\lambda=\sqrt{ [e^{-2\delta(r_\gamma)}\mu(r_\gamma)]
\frac{ N'(r_\gamma)}{2r_\gamma}}.\label{lambda2}
\end{equation}
From (\ref{NGR}) we have 
\begin{equation}
\begin{split}
N'(r_\gamma)&=(n-1)\mu'(r_\gamma)-\frac{2\kappa}{(n-2)}\left[2r_\gamma p_r(r_\gamma)+
r^2_\gamma p'_r(r_\gamma)\right]\\
&\geq (n-1)\mu'(r_\gamma)+2{\kappa} r_\gamma p_r(r_\gamma),\label{Npf}
\end{split}
\end{equation}
where in order to establish the last inequality, we have taken into account that 
at $r_\gamma$, $p_r$ and $P'$
are both nonpositive numbers.
However, from the first factor in (\ref{lambda2}), which is the factor 
$e^{-2\delta(r_\gamma)}\mu(r_\gamma)$, 
we  can only ensure that $e^{-2\delta(r_\gamma)}\mu(r_\gamma)\leq\mu(r_\gamma)$, 
and therefore
this inequality and (\ref{Npf}) impose bounds in opposite 
directions, thereby preventing 
a universal bound (independent of the matter content) 
from being written 
for the Lyapunov exponent.

Coming back to the inequality (\ref{omegacot}), it can also be used to 
impose a universal bound over the location of the first
relativistic image in the strong lensing regime. In general, if we assume 
that the observer
is at a distance $D_{ol}$ from the lens, then the first relativistic 
image subtends an 
angle $\theta_\infty$ \cite{Bozza02,Stefanov10}, which
can be expressed in terms of the circular orbital frequency $\Omega_\infty$ as
\begin{equation}
\theta_\infty=\frac{1}{D_{ol}\Omega_\infty}.
\end{equation}
From this relation and using (\ref{omegacot}),
we obtain the universal and simple lower bound
\begin{equation}
\theta_\infty\geq \frac{1}{D_{ol}}\left[\frac{r^2_H}{\frac{n-5}{(n-1)r^2_H}\hat\alpha+\frac{n-3}{n-1}}
\right]^{1/2}.
\end{equation}
In the GR limit it reduces to
\begin{equation}
\theta_\infty\geq \frac{\, r_H}{D_{ol}}\left[{\frac{n-1}{n-3}}
\right]^{1/2},
\end{equation}
which, for the 4-dimensional case gives
\begin{equation}
\theta_\infty\geq \sqrt{3}\frac{\, r_H}{D_{ol}}.
\end{equation}
\section{General comment on the fastest way to circle 
axially-symmetric spacetimes}
It would be very interesting to find a generalization of some 
of the previous results
for the case of axial symmetry. Even in this case, the 
circular null geodesics on the
equatorial plane share some of the properties found 
in the spherical symmetric case. 
For example, 
Wei and Liu (\cite{Wei14}) found universal relations between 
the first relativistic image and the quasinormal frequencies by 
analyzing equatorial circular null 
geodesics in arbitrary axially-symmetric black holes, thus 
extending some of the 
results of spherical symmetry.

 Recently, Hod\cite{Hod11} also showed that the fastest way to circle a 
black hole in general 
relativity is through circular null geodesics. In particular, 
he demonstrated 
that for any spherically 
symmetric spacetime in four dimensions, the minimum orbital 
time as measured by an 
asymptotic observer is realized by circular null geodesics, 
independently of the underlying gravitational theory. Moreover, 
he showed that a similar 
result can be obtained in GR for the case of 
a Kerr black hole by noting that the equatorial circular 
orbit with minimum traveling
time coincides with the solution obtained by solving the equation governing 
circular null geodesics.
More recently, in \cite{Pradhan13}, Pradhan, in his study of charged 
Myers-Perry black holes 
in higher dimensions, made the observation that Hod's 
conclusion regarding the fastest way 
to circle a black hole 
remains valid in this more general family of metrics.

As mentioned, the fact that circular null geodesics minimize 
the orbital period in 
the case of spherically symmetric spacetimes is always valid. 
However, in\cite{Hod11}, when Hod analyzed the Kerr metric,  
he obtained two different equations (one for the 
fastest circular orbit, and another for the 
circular
null geodesics, see Eqs.(21) and (31) in his paper), and
he showed that they admit the same kind of solutions; therefore, 
it was not clear from his 
discussion how general these results were. In other words, 
one can ask whether the agreement 
between equatorial 
circular null geodesics and the fastest way to circle
black holes is a property unique to GR or it is a 
geometrical property 
general to any axially-symmetric 
metric, independently of any gravity theory.

Additionally, Hod made the interesting conjecture that
there should be a lower bound for the traveling time, given 
in terms of the ADM mass 
of the black hole\cite{Hod11}. 
More precisely, he conjectured a lower bound on the orbital periods 
$T_\infty$ of circular null geodesics around compact objects 
with mass $\mathcal{M}$ 
(as seen from far away 
observers). In mathematical terms he states that (in units with $G=c=1$)
\begin{equation}
T_\infty\geq 4\pi \mathcal{M}\label{Tinf}.
\end{equation}
In particular, the equality is satisfied by an extreme Kerr black hole.
This conjecture is physically motivated by taking into 
account the rotational dragging 
which is maximal in the case of an extremal black hole.
If this conjecture were correct, it could also be used to 
establish another universal bound for
the possible location of the first relativistic image of an 
object lensed by a black hole.
In fact, if (\ref{Tinf}) were valid we should obtain 
$\Omega_\infty=2\pi/T_\infty\leq\frac{1}{2\mathcal{M}},$ which would imply
\begin{equation}
\theta_\infty\geq\frac{2\mathcal{M}}{D_{ol}}.\label{thetab}
\end{equation}
In other words, the expression (\ref{thetab}) would also be valid
for a generic rotating black hole, at least in the GR regime.

Even given the reasonableness of conjecture (\ref{Tinf}), 
it has not yet been proven. 
However, it leads one to wonder whether this conjecture 
can be kept as potentially valid in alternative 
gravitational theories. 
It is our purpose in this section to answer these questions. 
In particular, we 
make the observation 
that the orbital period for circular orbits {\it always} 
coincides with the  circular null geodesics for 
any axially-symmetric spacetime, 
independently of the gravitational theory.
We also answer the second question negatively 
on the validity of (\ref{Tinf}) for alternative
black hole candidates 
by giving an explicit counterexample where the bound 
assumed by Hod is not satisfied. 
What is more, we will show that the conjecture can be violated 
even in the case of a nonrotating 
black hole.
In order to do that, we will discuss a special family of 
Kaluza-Klein black holes.

Let us start with the first question.
We begin, with the more general conformally stationary and
axially-symmetric metric in $n$-dimensions, without making 
reference to any gravitational theory.

These kinds of metrics are given by
\begin{equation}
\begin{split}
ds^2&=e^{2\Phi(r,\theta,t)}\left(g_{tt}dt^2+2g_{t\phi}dtd\phi+g_{rr}dr^2\right.
\\&+\left. g_{\theta\theta}d\theta^2+g_{\phi\phi}d\phi^2+r^2\cos(\theta)^2d\Omega^2_{n-3}\right),
\end{split}
\end{equation}
with
\begin{equation}
d\Omega^2_{n-3}=d\phi^2_2+\sin^2\phi_2\left[d\phi^2_3+\sin^2\phi_3(\cdots d\phi^2_{n-4})\right],
\end{equation}
and with the components of the metric not depending on $\phi$.
As this metric is axially-symmetric, there are equatorial orbits, 
and in particular we
can compute from this metric the orbital time for circular null curves, 
i.e., curves 
such that $ds^2=0, \theta=\pi/2$ and $r=r_\gamma$. 
The orbital period for these curves is:
\begin{equation}\label{eq:T}
T_\infty=2\pi\frac{-g_{t\phi}\pm\sqrt{\Delta}}{g_{tt}},
\end{equation}
where 
\begin{equation}
\Delta=g_{t\phi}^2-g_{tt}g_{\phi\phi},
\end{equation}
and the $+/-$ signs correspond to counter-rotating/co-rotating orbits, 
respectively.
If we compute their extremes we obtain
\begin{equation}
\frac{dT}{dr}=\frac{\pi}{\sqrt{\Delta}}G=0,
\end{equation}
with 
\begin{equation}\label{eq:G}
\begin{split}
G_\pm&=\frac{1}{g_{tt}^2}\left[(\mp 2g_{t\phi}\sqrt{\Delta}
+\Delta+g_{t\phi}^2)\frac{dg_{tt}}{dr}\right.\\
&\left.+(-2g_{tt}g_{t\phi}\pm 2g_{tt}
\sqrt{\Delta})\frac{dg_{t\phi}}{dr}+g_{tt}^2\frac{dg_{\phi\phi}}{dr}\right].
\end{split}
\end{equation}
So, if we assume that $\Delta$ is a regular function and we 
only consider the region exterior
to all horizons (where $\Delta=0$), the minimum orbital 
period is obtained for 
those curves which satisfy $G_\pm=0$.

On the other hand, we can study the condition for the 
existence of circular null geodesics on the
equatorial plane.
We start from the conservation equations
\begin{eqnarray}
E&=&-\left(g_{tt}\dot{t}+g_{t\phi}\dot{\phi}\right),\\
L&=&g_{t\phi}\dot{t}+g_{\phi\phi}\dot{\phi},\\
p_r&=&g_{rr}\dot{r},
\end{eqnarray}
which represent the energy of the null particle as measured by an 
asymptotic observer,
the orbital angular momentum and the radial component of the linear 
momentum, respectively.
These equations can be solved for $\dot{t}$, and $\dot\phi$ and replaced in
the Hamiltonian $H$,
\begin{equation}
H=-E\dot{t}+L\dot\phi+p_r\dot{r}=0;
\end{equation} 
obtaining the following equation for $\dot{r}$:
\begin{equation}
\dot{r}^2=V_r;
\end{equation}
with 
\begin{equation}
V_r=-\frac{E^2}{g_{rr}\Delta}\left(g_{\phi\phi}+2g_{t\phi}b
+g_{tt}b^2\right),
\end{equation}
and $b=L/E$, the impact parameter. 
Circular null geodesics must satisfy $\dot{r}_\gamma=\dot{r}'_\gamma=0$, or equivalently 
$V_r=V'_r=0$.
From  $V_r=0$, we obtain
\begin{equation}
b=\frac{\pm\sqrt{\Delta}-g_{t\phi}}{g_{tt}},\label{eq:b}
\end{equation}
and from $V'_r=0$ it follows that
\begin{equation}
\frac{d}{dr}g_{\phi\phi}+2b\frac{d}{dr}g_{t\phi}+b^2\frac{d}{dr}g_{tt}=0,
\end{equation}
Finally, by replacing the expression for $b$ 
given by (\ref{eq:b}) in the last equation,
we conclude that the circular null geodesics, if they exist, 
must satisfy
\begin{equation}
\begin{split}
&\frac{1}{g_{tt}^2}\left[(\mp 2g_{t\phi}\sqrt{\Delta}
+\Delta+g_{t\phi}^2)\frac{dg_{tt}}{dr}\right.
\\&\left.+(-2g_{tt}g_{t\phi}\pm 2g_{tt}\sqrt{\Delta})
\frac{dg_{t\phi}}{dr}+g_{tt}^2\frac{dg_{\phi\phi}}{dr}\right]=0,
\end{split}
\end{equation}
which agrees with the expression for $G_\pm$ given previously.
As a result, its critical points always coincide with 
the values of the circular geodesics, generalizing in this way 
the results given by Hod for
spherically symmetric spacetimes, and the particular cases of Kerr\cite{Hod11} 
and Myers Perry\cite{Pradhan13}. Let us remark that 
there is not inconsistency 
between this result
and the fact that in\cite{Hod11} Hod obtained
two different algebraic equations [Eqs.(21) and (31) in \cite{Hod11} for 
the minimum orbital period  
and geodesic motion, respectively]. In fact, 
the circular null geodesics are obtained by solving $\dot{r}_\gamma=0$ 
and  $\dot{r}'_\gamma=0$, 
both being functions of $b$. If we first solve $\dot{r}'_\gamma=0$ for $b$ 
and replace its value in 
$\dot{r}_\gamma=0$, (as the path followed by Hod) we obtain Eq.(31) 
of\cite{Hod11}. Alternatively, 
if we first solve for $b$ 
from $\dot{r}_\gamma=0$ and replace it in $\dot{r}'_\gamma=0$, 
we arrive to his Eq.(21).   

As mentioned above, in \cite{Hod11}, by comparing the total 
orbital period for rotating Kerr
black holes with those of Schwarzschild ones, 
a lower bound for the orbital 
periods 
$T_\infty$ (as measured from far away 
observers) of circular null geodesics around compact objects with 
mass $\mathcal{M}$, 
Eq.(\ref{Tinf}) was conjectured. 
Now, by giving an explicit counterexample, we will show that, 
in contrast to the geometrical 
status
of fastest way to circle black holes, this conjecture cannot be 
generally valid in other black holes candidates coming from 
alternative gravitational theories.

The counterexample is based on the Kaluza-Klein 
black hole\cite{Aliev13}. Shadows of these kinds of 
black holes were recently analyzed in\cite{Eiroa13}.
Let us consider the following metric, which satisfies 
the vacuum Einstein equations in five dimensions. 
\begin{equation}
ds^2=-(1-\frac{2m}{r})dt^2+\frac{dr^2}{1-\frac{2m}{r}}
+r^2(d\theta^2+\sin^2(\theta)d\phi^2)+dy^2.
\end{equation}
Then, by doing a compactification of the extra dimension 
and a boost transformation with 
velocity $v$ in the $y$-direction, 
and by projecting in the $4$-manifold, a new metric is obtained 
representing a charged 
spherically symmetric black hole together with a 
dilaton field\cite{Aliev13}. 
The four-dimensional metric reads:
\begin{equation}
\begin{split}
ds^2&=-\frac{1-\frac{2m}{r}}{B}dt^2+\frac{B}{1-\frac{2m}{r}}dr^2+\\
& Br^2\left(d\theta^2+\sin(\theta)^2d\phi^2\right),
\end{split}
\end{equation}
with
\begin{equation}
B=\sqrt{1+\frac{2mv^2}{r(1-v^2)}}.
\end{equation}
This metric together with the
scalar field 
\begin{equation}
\Phi=-\frac{\sqrt{3}}{2}\ln{B},
\end{equation}
and the electromagnetic potential $\mathcal{A}$
\begin{equation}
\mathcal{A}=\frac{v}{2(1-v^2)}\frac{1-\frac{2m}{r}}{B^2}dt,
\end{equation}
satisfies the set of Einstein-Maxwell-dilaton field equations 
that follow from the action:
\begin{equation}
S=\int d^4x\sqrt{-g}\left[R+2(\nabla\Phi)^2
-e^{2\sqrt{3}\Phi}F^2\right].
\end{equation}
This metric represents a black hole with ADM mass $\mathcal{M}$ and
charge $Q$ given in terms of the parameter $m$ and the velocity $v$ 
by:
\begin{eqnarray}
\mathcal{M}&=&m\left[1+\frac{v^2}{2(1-v^2)}\right],\\
Q&=&\frac{mv}{1-v^2}.
\end{eqnarray}

One of the characteristics of this type of metric is that the ADM mass,
the dilation field and the charge depend on the boost parameter.
Note that the event horizon is located at $r_H=2m$, which 
for a fixed mass $\mathcal{M}$, shrinks 
to zero when $v$ goes to $1$.
Let us compute now the minimum orbital time that a circular null geodesic 
takes to orbit this black hole.
In order to do this, we must solve for the co-rotating orbits, that is $G_{-}=0$, 
which, using the expression (\ref{eq:G}) in this case gives
\begin{equation}
\begin{split}
&(v^2-2)^2r^2-2\mathcal{M}(v^2-2)(4v^2-3)r\\
&-16\mathcal{M}^2v^2(1-v^2)=0.
\end{split}
\end{equation}
The physical solution of this equation is
\begin{equation}
r_\gamma=\frac{3-4v^2+\sqrt{9-8v^2}}{2-v^2}\mathcal{M}.
\end{equation}
By replacing this radius in (\ref{eq:T}), 
we obtain
\begin{equation}
T_\infty=2\pi\frac{3-4{v}^{2}+\sqrt{-8{v}^{2}+9}}{2-{v}^{2}}\sqrt{\frac{
3+\sqrt{-8{v}^{2}+9}}{-1+\sqrt{-8{v}^{2}+9}}}\mathcal{M}.
\end{equation}
A plot of this function is shown in Fig. 2. At $v=0$ (Schwarzschild case)
it takes the value $T_\infty=6\sqrt{3}\pi\mathcal{M}$, 
and it goes to zero when $v\rightarrow 1$.

\begin{figure}[]
\centering
\includegraphics[clip,width=80mm]{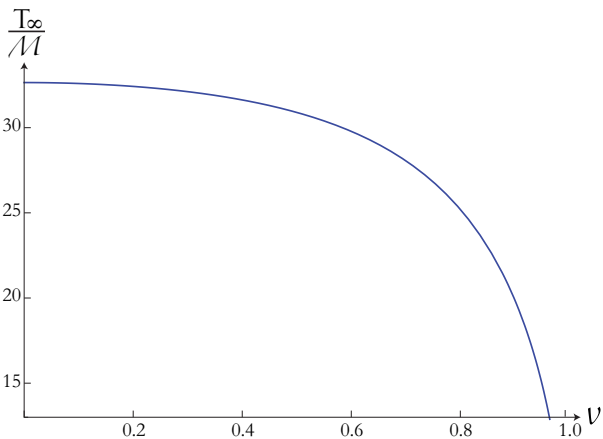}
\caption{Plot of the quotient $T_\infty/\mathcal{M}$ as 
a function of $v$ We assume $G=1$.
}
\label{fig:Tinfi}
\end{figure}
Consequently, the orbital period around a black hole with mass $\mathcal{M}$
can be made arbitrarily small.
Therefore, there is a value of $v$ from which the Hod conjecture 
cannot be satisfied for this type of black holes. 
Numerically we found that for $v>0.96591$ all the orbital periods $T_\gamma$ 
are smaller than $4\pi\mathcal{M}$.  

\section{Summary}
In this work, we have found upper bounds for the radius of 
photon spheres in $n$-dimensional
GR and EGB theories, 
thereby generalizing the previous work of Hod in\cite{Hod13}.
In the general situation of a black hole dressed with matter 
fields, we have seen that the photon
sphere radius is always smaller than that corresponding 
to a vacuum black hole with the same mass. 
It would be interesting to know how generic these results 
are, in the sense of
how much they depend on gravitational 
field equations. In particular, it would be natural to study whether these results 
hold for the more general 
Lovelock gravitational theory or if they can be extended to other
theories like $f(R)$ ones.

With respect to the study of circular null geodesics in axial 
symmetry, we have observed that equatorial circular null geodesics 
always minimize 
the orbital time around a black hole. It is a geometrical result. Given 
the astrophysical importance of this type of geodesics, it would 
also be interesting to attempt a proof of the Hod conjecture (\ref{Tinf}). 
We wish to deal with these and other associated problems in the near future.\\

\section*{Acknowledgments}
E. Gallo thanks the Instituto de F\'isica y
Astronom\'ia, Universidad de Valpara\'iso for their kind hospitality, 
while working on this paper. 
E. Gallo acknowledge financial support from CONICET and SeCyT-UNC.
J. R. Villanueva is supported by Comisi\'on Nacional de Investigaci\'on Cient\'ifica 
y Tecnol\'ogica through FONDECYT grants No  11130695.
In this post-publisher version two typos were corrected: the signature of the metric in eq.(1) and a factor in eq.(102). We thanks Gary Gibbons and Chris Pope for bringing to our attention these typos. 

\begingroup\raggedright\endgroup

\end{document}